\documentclass[journal]{IEEEtran}

\pdfoutput=1

\usepackage{graphicx}
\usepackage{color}
\usepackage{latexsym}
\usepackage{amsmath}
\usepackage{amsthm}
\usepackage{amssymb}
\usepackage{bm}
\usepackage{indentfirst}
\usepackage{epsf}
\usepackage{epsfig}
\usepackage{graphicx}
\usepackage{epstopdf}
\usepackage{cite}
\usepackage{hyperref}
\usepackage{cleveref}
\usepackage{enumerate}
\usepackage{enumitem}
\usepackage{textcomp}

\graphicspath{{Fig/}}

\begin{document}

\title{Statistical-Based Privacy-Preserving Scheme with Malicious Consumers Identification for Smart~Grid }

\author{Alireza Ahadipour, Mojtaba Mohammadi, and Alireza Keshavarz-Haddad
\thanks{Alireza Ahadipour, Mojtaba Mohammadi, and Alireza Keshavarz-Haddad are with the School of Electrical and Computer Engineering, Shiraz~University, Shiraz, Iran. Email: \{ahadipour.alireza, mojtaba.mohammadi, and keshavarz\}@shirazu.ac.ir}
}

\maketitle

\begin{abstract}
As smart grids are getting popular and being widely implemented, preserving the privacy of consumers is becoming more substantial. Power generation and pricing in smart grids depends on the continuously gathered information from the consumers. However, having access to the data relevant to the electricity consumption of each individual consumer is in conflict with its privacy. One common approach for preserving privacy is to aggregate data of different consumers and to use their smart-meters for calculating the bills. But in this approach, malicious consumers who send erroneous data to take advantage or disrupt smart grid cannot be identified. In this paper, we propose a new statistical-based scheme for data gathering and billing in which the privacy of consumers is preserved, and at the same time, if any consumer with erroneous data can be detected. Our simulation results verify these matters.
\end{abstract}

\begin{IEEEkeywords}
Smart grid, Supplier, Data aggregator, Privacy, Statistical, Correlation coefficient.
\end{IEEEkeywords}

\IEEEpeerreviewmaketitle

\section{Introduction}
\label{sec:introduction}
\IEEEPARstart{R}{ecently}, the traditional power grids underwent an alteration to smart grids which leads to several benefits including enhanced reliability and resilience, higher intelligence and optimized control, decentralized operation, higher operational efficiency, more efficient demand management, better power quality, and fraud detection \cite{fadel2015survey}. The smart grid is envisaged to be the next generation of the traditional grid. In smart grid, the consumers minimize their expenses while providers maximize their revenue, hence, a win-win partnership can be achieved.

On the contrary to the traditional grids, there is a bidirectional information flow between suppliers and consumers in smart grids rather than a centralized unidirectional system. This feature enables the supplier to generate the electricity based on the demand; and at the same time, the supplier can define dynamic billing tariff, and regard to these tariffs that are sent to consumer periodically (e.g. every $15$ minutes), each consumer decides whether to decrease or increase its power consumption. Thus, electricity is consumed in a more efficient manner.
In the other direction of information flow in smart grids, consumers declare their need for electricity or their power consumption; indeed, consumers send their momentary electricity usage to the suppliers. As a result, unlike traditional grids, in smart grids suppliers provide electricity based on the demands of consumers to avoid wasting power \cite{alabdulatif2017privacy}.

In order to provide two-way communication in smart grid, the consumers should be equipped with smart meters by which they can measure their usage and send/receive messages over communication links such as  power-line, cable, fibre, or radio.

The classic approach for billing is to gather all power consumption information in a center, i.e., the consumers send their electricity usage to a server – which is responsible for gathering data – periodically by means of smart meters and then, the server dispatches the gathered information to a local or central database. The electricity bill for each consumer is calculated based on records of the consumers in the database. Criticism to this scheme is that the privacy is not preserved. As each individual consumer sends its usage, the pattern of its power consumption is apparent to the center; for instance, inhabitant’s personal schedules, habits, religion, and so on \cite{li2014enabling,lu2016privacy}.

Another trivial approach for billing is that the supplier sends time-varying tariffs periodically to the consumers and their smart-meters compute the electricity consumption price over a defined period (e.g. one month) based on the received tariffs. Eventually, at the end of each period, every consumer only sends its total billing amount to the supplier. In this case, the privacy of each consumer would be preserved; however, supplier cannot verify consumers' billing reports.
Consequently, some malicious consumers would take advantage or distrust smart grid by sending incorrect information on their power usages. In this case, historical analysis of electricity usage reports would not be useful for identifying malicious users.
For instance,  if the power consumption pattern of a consumer alters over time, this consumer would be considered as a consumer who is declaring incorrect information; while a malicious consumer who ever sends artificial data cannot be easily identified \cite{XiaXLZ2018}.

According to the aforementioned scenarios, the main challenge in communications between consumers and suppliers is preserving the privacy of consumers while identifying any malicious consumer in the smart grid. To address this challenge, we propose a new scheme called statistical-based privacy preserving (SBPP). An earlier version of SBPP was presented in \cite{sbpp1}. Notably, the present work completes our previous paper and provides more technical details and adds some new ideas for data gathering and fraud detection.

The proposed scheme enables privacy preserving data gathering and detecting the malicious consumers possible at the same time. SBPP exhibits an efficient solution for privacy preserving in terms of computation complexity and communication overhead. The key idea in SBPP scheme is to combine usages of different consumers at local data aggregators (to preserve privacy) and also sending accurate usage of some randomly selected consumers to the supplier. Then the supplier uses the data of accurate usage of different consumers over random periods of time to detect malicious consumers.  Simulation results verify that SBPP is reliable for detecting malicious consumers in different sabotaging scenarios.

The remainder of this paper is organized as follows: In Section~\ref{sec:relatedworks}, we briefly discuss related works. In Section~\ref{sec:systemmodel}, the system model is introduced. In Section~\ref{sec:proposedscheme}, we describe our proposed statistical-based scheme for data gathering in smart grid. In section~\ref{sec:simulation}, the simulation results of the proposed scheme are presented. Finally, we conclude the paper in Section~\ref{sec:conclusion}.

\section{Related Works}
\label{sec:relatedworks}
Recently, many researchers have paid attention to privacy-preserving solutions for smart grids. In this section, we briefly review some proposed schemes for privacy-preserved data gathering in smart grids.

In \cite{wong2014privacy} authors propose an algorithm for data collection with self-awareness protection. The paper considers some data aggregators and consumers in a smart grid where some of the respondents may not participate in contributing their personal data or submit erroneous data. To overcome this issue a self-awareness protocol is proposed to enhance trust of the respondents when sending their personal data to the data aggregators. In this scheme, all consumers collaborate with each other to preserve the privacy. They have hired an idea, which allows respondents to know protection level before the data submission process is initiated. The work is motivated by \cite{domingo2010coprivacy} and \cite{ferrer2011coprivacy}. In \cite{domingo2010coprivacy}, co-privacy (co-operative privacy) is introduced. Co-privacy claims that best solution to achieve privacy is to help other parties to achieve their privacy.

Many papers study self-oriented privacy protection methods. For example, \cite{golle2008data} introduces self-enforcing privacy (SEP) for e-polling. In SEP scheme, supplier allows the consumers to track their submitted data in order to protect their privacy. In this case, the consumers can accuse the supplier based on data they gathered during the collection process. Following this idea, a fair approach for accusation is presented in \cite{stegelmann2010towards}.

In \cite{kumar2010freedom}, a respondent-defined privacy protection (RDPP) is introduced. It means that respondents are allowed to determine their required privacy protection level before delivering data to data collector. Unlike other methods in which data aggregators decide about the privacy protection level, in this scheme the consumers can freely define the privacy protection level.

There are also some other researches on privacy-preserving data collection. For instance, in \cite{wang2016balanced} authors design a balanced anonymity and traceability for outsourcing small-scale linear data aggregation (called BAT-LA) in smart grid. Anonymity means that consumers’ identity should be kept secret and traceability means that imposter consumers should be traced. Here an important challenge is that many devices are not capable of handling required complicated computations. Hence, they have hired the idea of outsourcing computations with the help of public cloud. The paper utilizes elliptic curve cryptography and proxy re-encryption to make BAT-LA secure. BAT-LA is evaluated by comparing it with two other schemes, RVK \cite{wang2015tpp}, and LMO \cite{rottondi2013distributed} and it is shown that BAT-LA is more efficient in terms of confidentiality. 

The papers \cite{wang2016balanced} and \cite{chun2018privacy} focus on outsourcing to clouds or distributed systems. For encryption process, it is important to use a secure key management scheme. The cryptographic technique ensures that no privacy sensitive information would be revealed. But, there is still the challenge of how to efficiently query encrypted multidimensional metering data stored in an untrusted heterogeneous distributed system environment. \cite{jiang2018achieving} addresses this issue and introduces a high performance and privacy-preserving query (P2Q) scheme which provides confidentiality and privacy in a semi-trusted environment.

To obtain privacy of residential consumers, a scheme named APED is proposed in \cite{sun2013aped}. The paper employs a pairwise private stream aggregation. The scheme achieves privacy preserving aggregation and also executes error detection when some nodes fail to function normally. DG-APED is an improved form of APED, suggested in \cite{shi2015diverse}. DG-APED propounds diverse grouping-based protocol with error detection. This research added differential privacy technique to APED. Moreover, DG-APED has an advantage of being efficient in terms of communication and computation overhead compared to APED.

In \cite{jia2014human}, the authors present a new kind of attack, which adversary extracts information about the presence or absence of a consumer to access the smart meter information. The attack is called human-factor-aware differential aggregation (HDA) and it is claimed that other proposed schemes cannot handle it. To solve this issue, the authors introduce two privacy-preserving schemes which can stand out against HDA attack by transmitting encrypted measurements to an aggregator in a way that aggregator cannot steal any information of human activities.

PDA is a privacy-preserving dual-functional aggregation scheme in which, every consumer disseminates only one data and then supplier computes two statistical averages (mean and variance) of all consumers \cite{li2015pda}. The paper shows by simulations that PDA possesses low computational complexity and communication overheads. In another work, the authors introduce privacy-preserving data aggregation with fault-tolerance called PDAFT \cite{chen2015pdaft}. If PDAFT is implemented, a strong adversary is not able to gain any information, even in the case of compromising a few servers at the supplier. Like PDA, PDAFT has relatively high communication overhead and is tenacious against many security threats. In PDAFT scheme, if some consumers or servers fail, it can still work correctly.

DPAFT \cite{bao2015new} is another privacy-preserving data collection scheme which supports both differential privacy and fault tolerance at the same time. It is claimed that, DPAFT surpass other schemes in many aspects, such as storage cost, computation complexity, utility of differential privacy, robustness of fault tolerance, and the efficiency of consumer addition or removal \cite{bao2015new}. A new malfunctioning data aggregation scheme, named MuDA, is introduced in \cite{chen2015muda}. The scheme is resistant to differential attacks and keeps consumers’ information secret with an acceptable noise rate. PDAFT \cite{chen2015muda}, DPAFT \cite{bao2015new}, and MuDA \cite{chen2015muda}, shows nearly same characteristics with differences on their cryptographic methods \cite{ferrag1611survey}. PDAFT employs homomorphic Paillier cryptosystem \cite{paillier1999public}, while DPAFT and MUDA use Boneh-Goh-Nissim cryptosystem \cite{boneh2005evaluating}.

In \cite{fan2014privacy} authors present a secure power usage data aggregation method for smart grid where the supplier understands usage of each neighborhood and makes decision about energy distribution, while it has no idea of the individual electricity consumption of each consumer. This method is designed to barricade internal attacks and provide batch verification. Authors of \cite{he2016privacy} found out that \cite{fan2014privacy} has the weakness of key leakage and the imposter can obtain the private key of consumer easily. It is proved that by using the protocol in \cite{he2016privacy}, key leakage problem is solved and a better performance in terms of computational cost is achieved. Neglecting energy cost is the main disadvantage of this method.

In \cite{chun2018privacy}, a privacy-preserving protocol for smart grid is presented which outsources computations to cloud servers. In this protocol, the data is encrypted before outsourcing and consequently cloud can perform any computations without decryption. \cite{baloglu2018lightweight} adopts perturbation techniques to preserve privacy and uses perturbation techniques and cryptosystems at the same time. It is designed in a way to be suitable for hardware-limited devices. Evaluations show that it is resilient to two types of attack: filtering attack, and true value attack.

Authors of \cite{rial2018privacy} explain how for privacy preserving an individual meter of a consumer can share its readings to multiple consumers, and how a consumer can receive meter readings from multiple meters. They propose a polynomial-based protocol for pricing. In \cite{csimcsek2018tps3} a security protocol called TPS3 is introduced which uses Temporal Perturbation and Shamir’s Secret Sharing (SSS) to guarantees privacy and reliability of consumers’ data. In \cite{liao2017optimal}, data collector tries to preserve privacy by adding some random noise to its computation result. To overcome the problem of computation accuracy reduction, an approximation method is proposed in \cite{liao2017optimal} which leads to obtain a closed form of aggregator’s decision problem. In \cite{xu2018privacy}, a slightly different scenario is considered in which a data aggregator collects data from consumers and then spreads data to supplier. The goal is to preserve consumers’ data privacy. Anonymization might be an answer, but it has its own challenges. To achieve a tradeoff between privacy protection and data utility, interactions among three elements of scenario (consumers, data aggregator, and supplier) is modelled as a game and the Nash equilibria of the game is found.

In this paper we use the idea of aggregating data of different consumers for persevering the privacy of each individual consumer. The proposed SBPP scheme applies a simple statistical method for identifying the malicious consumers who send erroneous data to the aggregator in order to take advantage or disrupt smart grid. SBPP does not require any change on consumer side and communication infrastructure. It possesses very low computational overhead in aggregators and billing center in order to preserve privacy and detect malicious consumers. Therefore, SBPP is practical in the sense that it can be easily implemented on the existing infrastructure of smart grids.

\section{System Model}
 \label{sec:systemmodel}

In this section, we present our system model. The essential elements of our proposed scheme include:

\textit{Consumer:} those who consume energy in a power grid.

\textit{Benign consumer:} a consumer who reported its power consumption correctly.

\textit{Malicious consumer:} a consumer who reported its power consumption incorrectly due to some purposes such as fraud (power theft) or disruptive goals (power loss).

\textit{Supplier:} an entity whose responsibility is to provide energy for power consumers in a region.

\textit{Data aggregator:} a local facility whose liability is gathering the amount of power consumption information from consumers periodically and dispatching the gathered data to the supplier.

\textit{Electricity leakage}: the difference between the actual amount of consumed energy and the sum of quantity reported by consumers as their power consumption.

We consider a power grid consisting of $M$ regions where each region comprises one data aggregator. Denote the number of consumers in the $j$'th region by $n_j$ for $j= 1, \ldots, M$. Assume that the consumers send their power consumption information measured by the smart meters to the local aggregators. The aggregators are responsible of gathering local data and sending some information regarding the usage of the consumers to the power supplier.

It is assumed that data aggregators are trusted. Indeed, no information leakage occurs at data aggregators, supposedly because after aggregation takes place, no raw information concerning power consumption of consumers would be at hand. Besides, we assume that communications among the above entities of smart grid are secured. This means that the submitted data by a consumer cannot be altered in the communication infrastructure, i.e., the erroneous data would be only generated by a consumer itself.

\section{Proposed Scheme}
 \label{sec:proposedscheme}
Although the accuracy of smart grids' performance is engaged with the correctness of data gathered from consumers, this data gathering should not violate the privacy of consumers. Here we propose a new privacy-preserving data gathering scheme with the purpose of informing the supplier of the instant power consumption. The proposed scheme provides the power usage information for the supplier while keeping the consumers' power consumption information private and more importantly, finds malicious consumers in the process. The proposed scheme possesses very low computational complexity and communication overhead on the smart grid in comparison with the existing methods.

\subsection{Data Gathering in SBPP}
Here we present Statistical-Based Privacy-Preserving Scheme (SBPP) for data gathering in smart grids. SBPP gathers the information in the following steps:
\begin{enumerate}
	\item [1.] Consumers report their power consumption periodically to the local data aggregator.
	\item [2.] At each time period, the aggregator computes the total amount of power consumption based on the gathered data from the consumers. It also randomly selects the reported value of one the consumers.
	\item [3.] The aggregator sends the total power consumption value along with the name and reported value the selected consumer to the supplier.
	\item [5.] The supplier provides energy based on the reports of the aggregators and stores the reported values for randomly selected consumers.
\end{enumerate}

\begin{figure*}[t!]
	\begin{center}
		\includegraphics[width=\textwidth]{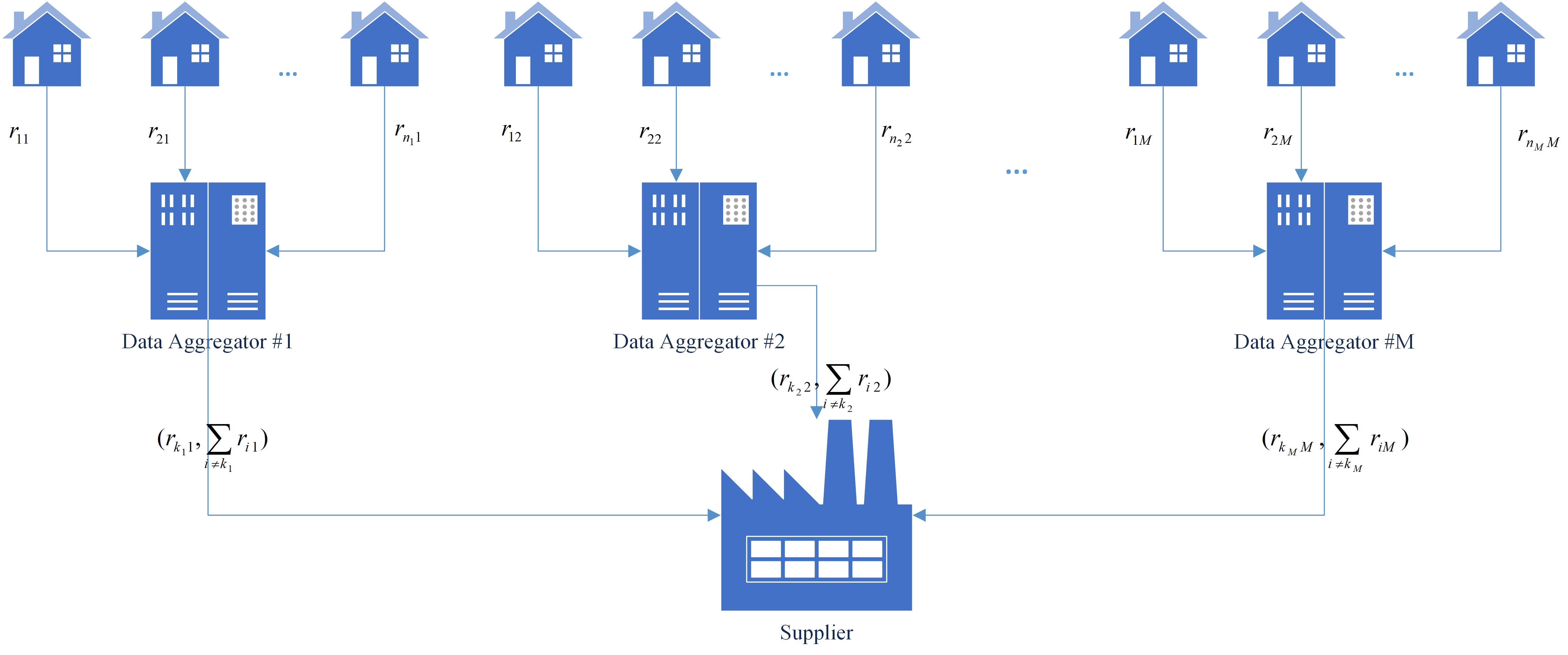}
	\end{center}
	\caption{The total power consumption is calculated by aggregators in each region and sent to the supplier. Let $r_{ij}$ be the reported consumed power by consumer $i$ in region $j$ and $k_j$ denotes the index of randomly chosen consumer in region $j$.}
	\label{datagathering}
\end{figure*}

Fig.~\ref{datagathering} depicts how data gathering takes place.

Here it is assumed that data aggregators are trusted and the power consumption data is not at hand any more after being summed up by the data aggregators. Based on this assumption, instead of having access to power consumption data of each individual consumer at any period of time, a little portion of information is available about the power consumption of each consumer.

As an example, suppose there are $100$ consumers in a region with one data aggregator and let the period of data gathering be $15$ minutes. Without this data gathering algorithm mechanism, a consumer should send its power consumption information to the supplier $30*24*60/15=2880$ times in a month. By employing our scheme for data gathering, on average $2880/100=28.8$ values in regards to the power consumption of each consumer is available at the supplier in an analogous period of time. Although it may seem that having access to power consumption information of consumers is in contradiction with their privacy, availability of these information $28$ times a month on random periods would not reveal any data concerning their life style compared with approachability of these information $2880$ times within a month.

\subsection{Detecting Malicious Consumers}
Malicious consumers pursue two distinct aims by sending erroneous data to suppliers. Either they declare their amount of power consumption lower than their real consumed power in order to pay less fee; or, they report their power consumption quantity much higher to impose more expenditure to the supplier.

In this paper, we get use of Pearson correlation coefficient of power consumption of consumers in order to find malicious consumers in each region who try to send erroneous data to the supplier.

Pearson correlation coefficient illustrates the statistical relationship between two variables and it is defined as follows:
\begin{equation}
corr(x,y) =  \, \frac{cov(x,y)}{\sqrt{var(x) \cdot var(y)}}
\label{corr}
\end{equation}
where $corr$ is a widely used alternative notation for the correlation coefficient and $cov$ means covariance.

Correlation coefficient possesses values in the range of $[-1,+1]$, where $+1$ and $-1$ indicate the strongest possible agreement and disagreement respectively.

In order to find malicious consumers, it is assumed that data aggregators are aware of the total amount of power consumed in each region. By comparing this amount with the aggregated quantity declared by consumers, the electricity leakage value (shortage) can be determined.

Having access to merely one quantity of power consumption information corresponding to a consumer does not suffice to distinguish if that consumer is benign or malicious. In other words, the more information we have regarding power consumption of each consumer, the better decision we can make about the sabotage of consumers. Thus, the scheme for finding malicious consumers can take place at the end of month or after a few months.

In order to detect malicious consumers, each data aggregator stores the identity (ID) of the randomly selected consumer, its reported power consumption ($r$), and the electricity leakage amount of power consumed in that region at every period ($l$). Then for each consumer, the data aggregator computes the correlation coefficient of its reported consumed energy and the leakage values of power consumption. The leakage quantity is defined as:
\begin{equation}
l = c - r
\label{leakage}
\end{equation}
where $c$ and $r$ are the actual and reported power consumption values respectively.

It is straightforward to see that the correlation coefficient turns to $0$ for benign consumers, since the leakage value is statistically independent from the power consumption of a benign consumer. Now if the correlation coefficient turns to $+1$ for a consumer, this means that (according to (\ref{leakage})) the consumer is reporting its power consumption less than its actual used power. On the other hand, if the correlation coefficient for a consumer turns to $-1$, it means that consumer is declaring its power consumption more than its usage due to some subversive goals.
Thus, SBPP scheme is capable of not only detecting malicious consumers, but also comprehending if that consumer is declaring its amount of power consumption less or more than its actual quantity.

Next, we study the performance of SPBB for some scenarios based on the number of malicious consumers in each region and the behavior of attackers.

\subsubsection{Scenario I: Existence of at most one malicious consumer in each region}
\

Suppose that there exists at most one malicious consumer in each region. According to the declared quantity of power consumption of consumers, three cases could be considered:

It should be stated that in all of the following formulas, all the variables are zero-mean. Indeed, the mean-value of all variables are subtracted from their real values. Throughout the paper, zero-mean vectors are depicted as $\bar{\textbf{x}}$.

\textit{\textbf{Case I:} Malicious consumer reports a portion/multiple of its actual usage:}
In this case, the reported quantities by the malicious consumer would undoubtedly have correlation with leakage amounts. Consider an arbitrary consumer, and let vectors ($\textbf{\underline{r}}$) and ($\textbf{\underline{l}}$) be the zero-mean vectors containing reported values of that consumer and the corresponding electricity leakage amounts on those periods. Here, the correlation coefficient for each consumer would be written as:
\begin{equation}
corr(\bar{\textbf{r}},\bar{\textbf{l}}) = \frac{\bar{\textbf{r}}^T \bar{\textbf{l}}}{\|\bar{\textbf{r}}\| \|\bar{\textbf{l}}\|} \\
\label{malcorr}
\end{equation}

In this case, the malicious consumer reports a portion/multiple of its consumed energy, i.e., $r = \alpha c$, where $\alpha$ is a positive coefficient, $\alpha > 0$. Thus, by getting use of (\ref{leakage}), for a malicious consumer, the correlation coefficient in (\ref{malcorr}) would be as:
\begin{align}
corr(\bar{\textbf{r}},\bar{\textbf{l}})
\nonumber
& =\frac{\bar{\textbf{r}}^T \bar{\textbf{l}}}{\|\bar{\textbf{r}}\| \|\bar{\textbf{l}}\|} \\
\nonumber
& = \frac{\alpha \bar{\textbf{c}}^T (1-\alpha) \bar{\textbf{c}}}{|\alpha| \|\bar{\textbf{c}}\| |1-\alpha| \|\bar{\textbf{c}}\|} \\
& = \frac{\alpha(1-\alpha)\|\bar{\textbf{c}}\|^2}{|\alpha||1-\alpha|\|\bar{\textbf{c}}\|^2} \\
\nonumber
& = \frac{\alpha(1-\alpha)}{|\alpha||1-\alpha|}
\label{case1}
\end{align}

As stated before, when the malicious consumer reports its power consumption less than its actual quantity, $0 < \alpha < 1$, the correlation coefficient turns to +1, and on the contrary, the correlation coefficient turns to -1 when the malicious consumer declares its consumed energy more than its actual usage, $\alpha > 1$.

\textit{\textbf{Case II:} Malicious consumer adds/subtracts a fixed quantity to/from its actual usage:}
In this case, the reported quantity by the malicious consumer is independent from the leakage amount until the actual consumed energy lies below the fixed value ($\eta$) which is added/subtracted to/from the actual power usage. Indeed, as the reported consumed energy cannot be negative, the reported quantity and the corresponding electricity leakage for each period could be written as:
\begin{align}
& r =
\begin{cases}
c-\eta & if \; c \geq \eta \\
0 & if \; c < \eta
\end{cases}\\
& l=
\begin{cases}
\eta & if \; c \geq \eta \\
c & if \; c < \eta
\end{cases}
\end{align}

While the actual consumed energy in each period is greater than the fixed threshold $\eta$, $c \geq \eta$, these terms become independent and thus, the malicious consumer cannot be detected. On the other hand, while the consumed power is less than the threshold, $c \leq \eta$, the malicious consumer would report its consumed energy zero and thus, the reported consumed power would be dependent to the leakage quantity. Consequently, by focusing on the measurements where $r$ is small, we can still detect the malicious consumer.

\textit{\textbf{Case III:} Malicious consumer adds/subtracts a random quantity to/from its actual usage:}
Assume that the malicious consumer adds/subtracts a random value independent from its power consumption to/from its consumed energy such that none of its reported quantities turns to a non-negative value. In this case, although the declared amounts of power consumption are quite independent from the electricity leakage corresponding to them at each period, the proposed scheme is capable of detecting the malicious consumer as well.
\begin{align}
corr(\bar{\textbf{r}},\bar{\textbf{l}})
\nonumber
& =\frac{\bar{\textbf{r}}^T \bar{\textbf{l}}}{\|\bar{\textbf{r}}\| \|\bar{\textbf{r}}\|} \\
\nonumber
& = \frac{(\bar{\textbf{c}} - \bar{\boldsymbol{\theta}})^T \bar{\boldsymbol{\theta}}}{\|\bar{\textbf{c}} - \bar{\boldsymbol{\theta}}\| \|\bar{\boldsymbol{\theta}}\|} \\
& = \frac{\bar{\textbf{c}}^T \bar{\boldsymbol{\theta}}}{\|\bar{\textbf{c}} - \bar{\boldsymbol{\theta}}\| \|\bar{\boldsymbol{\theta}}\|} -
\frac{\|\bar{\boldsymbol{\theta}}\|}{\|\bar{\textbf{c}} - \bar{\boldsymbol{\theta}}\|}
\label{case3}
\end{align}
where $\bar{\boldsymbol{\theta}}$ is a vector containing random values ($\theta$s) added/subtracted to/from the reported quantity. While $\bar{\textbf{c}}$ and $\bar{\boldsymbol{\theta}}$ are independent, the first term in (\ref{case3}) would be equal to zero and thus, the correlation coefficient corresponding to the malicious consumer turns to a negative quantity. As the correlation coefficient quantities corresponding to benign consumers revolve around $0$, the malicious consumer should take the smallest negative quantity amongst others.

\subsubsection{Scenario II: Existence of more than one malicious consumer in each region}
\

Furthermore, it is possible that there are more than one malicious consumer in a region. In this case, although the correlation coefficient corresponding to these consumers would not be equal to \textpm{1}, their correlation coefficient can be still distinguished from other consumers. As a result, it is needed that a threshold ($th$) must be defined where the absolute value of correlation coefficients fewer or more than the threshold indicate benign or malicious consumers respectively, as:
\begin{align}
 \begin{cases}
malicious \; consumer, & if \;\;\, -1 \leq corr \leq -th \\
benign \; consumer, & if \; -th < corr < th \\
malicious \; consumer, & if \;\;\;\;\;\, th \leq corr \leq 1
\end{cases}
\label{Threshold}
\end{align}

It is apparent that for higher threshold value, fewer malicious consumers are detected and on the other hand, lower threshold value, more benign consumers are considered as malicious consumers. Thus, a question that arises here is that how a proper threshold be found? The analysis concerning the detection of several malicious consumers with small number of samples for each consumer, is out of scope of this paper, however, we briefly discuss the problem in the next section. In this paper, according to the setting of the problem, we set the threshold to a fixed value, namely, $0.5$.

As the proposed scheme is a statistical technique, it is probable that the correlation coefficient of a benign consumer lies out of its defined region depicted in (\ref{Threshold}), or vice versa.

\subsection{Billing}
Here we describe billing procedure in SBPP scheme. The accountability for billing is handled by data aggregators. As discussed in the last section, malicious consumers can be identified by analyzing the correlation coefficient of each consumer in the region. Malicious consumers' being detected, the sent data corresponding to other consumers are considered trustworthy and error free. Based on this assumption, the liability for billing can be assigned to data aggregators. In every period, consumers send their amount of consumed energy to data aggregators. Then, based on the received data from the consumers and the received tariffs from the supplier, the data aggregators compute the cost of consumed power for each consumer before data aggregation takes place. In each period, data aggregators calculate the cost of consumed power for each consumer and add the cost to the previously calculated cost for that consumer and by the end of month, a bill will be issued and sent to the supplier's accounting center. Note that the task of computing the bills can be assigned to the smart-meters as well.

Not only this scheme decreases the signalling overhead, but also the privacy of consumers will be protected. It is merely required that suppliers send tariffs periodically to data aggregators and to consumers simultaneously. Data aggregators compute the cost of consuming energy for every consumer and smart meters on the consumers' side adjust the power consumption based on the received tariffs, i.e., when the tariff increases, smart meters force dispensable devices to be turned off. In this case, no information leakage and thus no privacy invasion would occur.

\section{Simulation Results}
\label{sec:simulation}
In this section, we describe our simulation results for the proposed SBPP scheme. The results verify that SBPP scheme is capable of detecting malicious consumers who send bogus information concerning their power consumption.

Although in reality there exists a dependency between power consumption of consumers in every successive periods, in all simulations we consider a random power consumption for each consumer in each period, which is the worst case that could be considered. We show that our proposed method works properly in this case and thus could be applied in real world smart grid.

Here, the previously described scenarios are investigated.

\subsection{Case I: Malicious consumer multiplies its usage}
\begin{figure}[t!]
	\begin{center}
		\includegraphics[width=\linewidth]{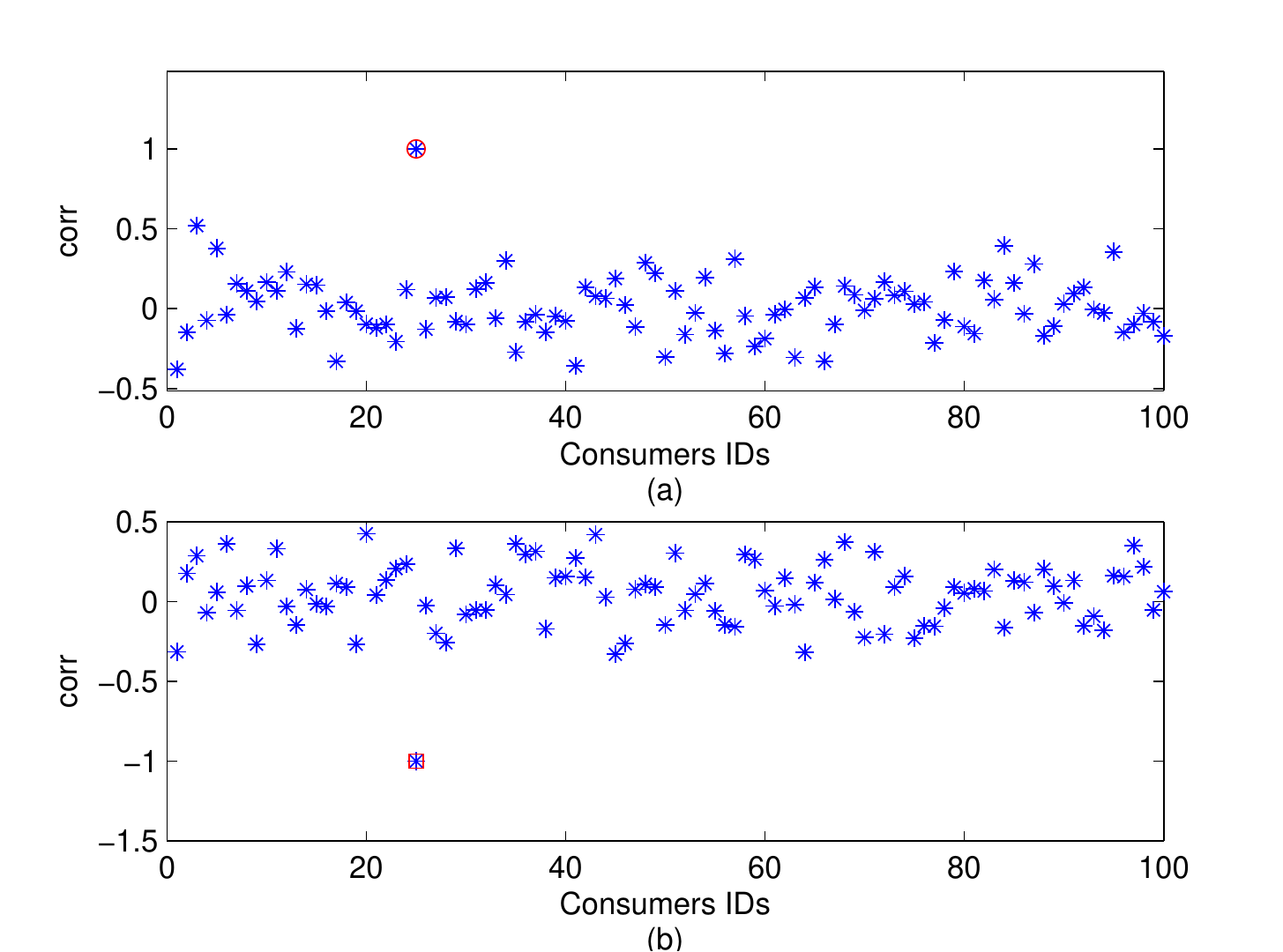}
		\caption{Correlation coefficient of the reported consumed energy and the leakage amounts of power consumption for all consumers in the grid. (a) The scenario where malicious consumer declare its power consumption less than the actual quantity and (b) the scenario where malicious consumer declare its power consumption more than the actual quantity}
		\label{Malicious}
	\end{center}
\end{figure}

In the following, all simulations are considered based on this assumption.
Consider a region consisting of $100$ consumers and one data aggregator where data aggregation takes place every $15$ minutes. Assume that the consumer $\# 25$ is a malicious consumer. Two cases are studied; consumer $\# 25$ in scenario (i) reports one tenth ($0.1$) of its power consumption and in scenario (ii) it declares its power usage $10$ times more than its actual consumption. Fig.~\ref{Malicious}~(a) illustrates scenario (i) where the correlation coefficient of reported consumed energy and the leakage amounts of power consumption turns to $+1$ and Fig.~\ref{Malicious}~(b) depicts scenario (i) where the correlation coefficient turns to $-1$.

Note that henceforth, all the assumptions are analogous to that of Fig.~\ref{Malicious} where consumer $\# 25$ is the malicious consumer unless mentioned otherwise.

\begin{figure}[t!]
	\begin{center}
		\includegraphics[width=\linewidth]{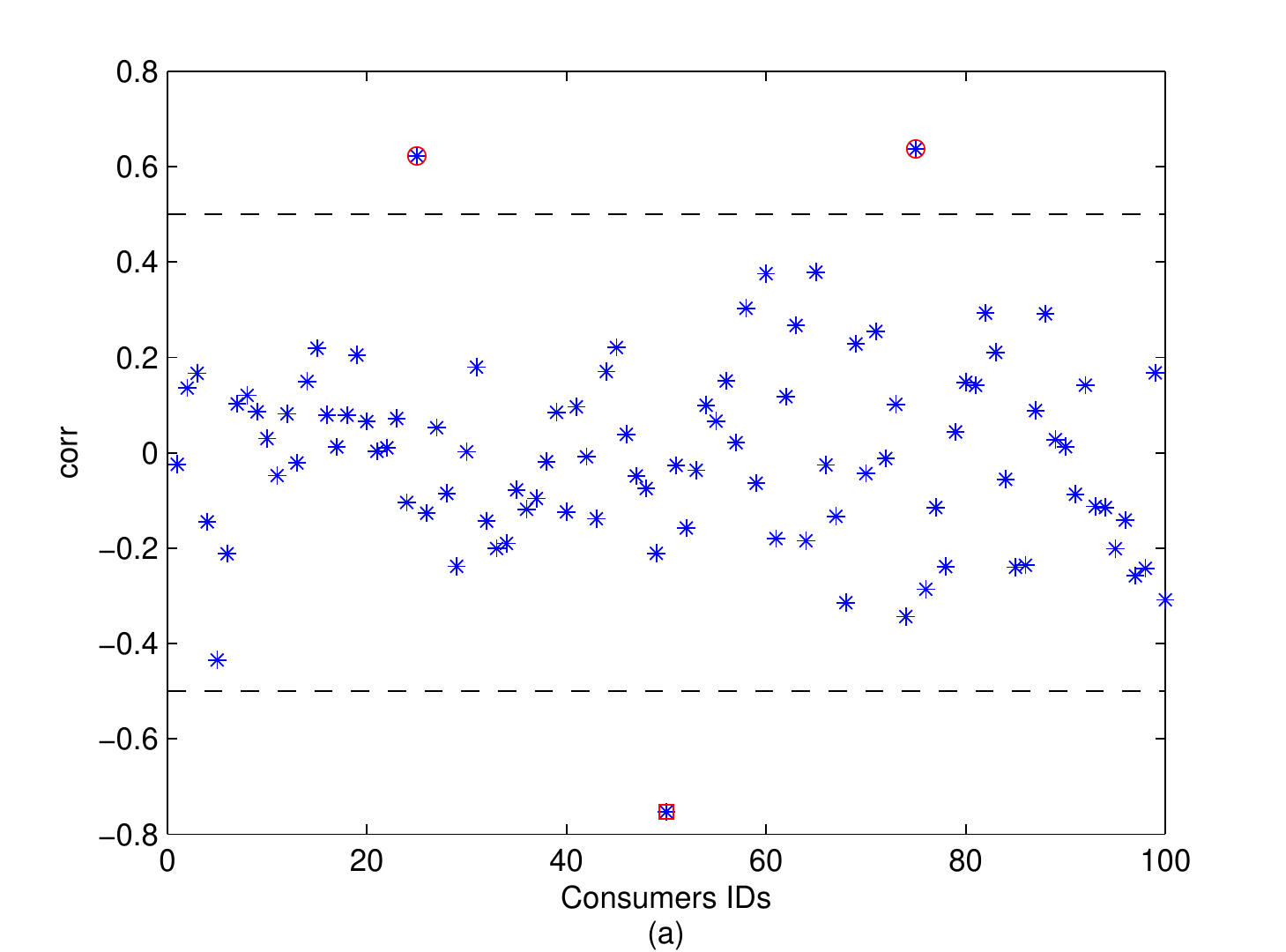}\\
		\includegraphics[width=\linewidth]{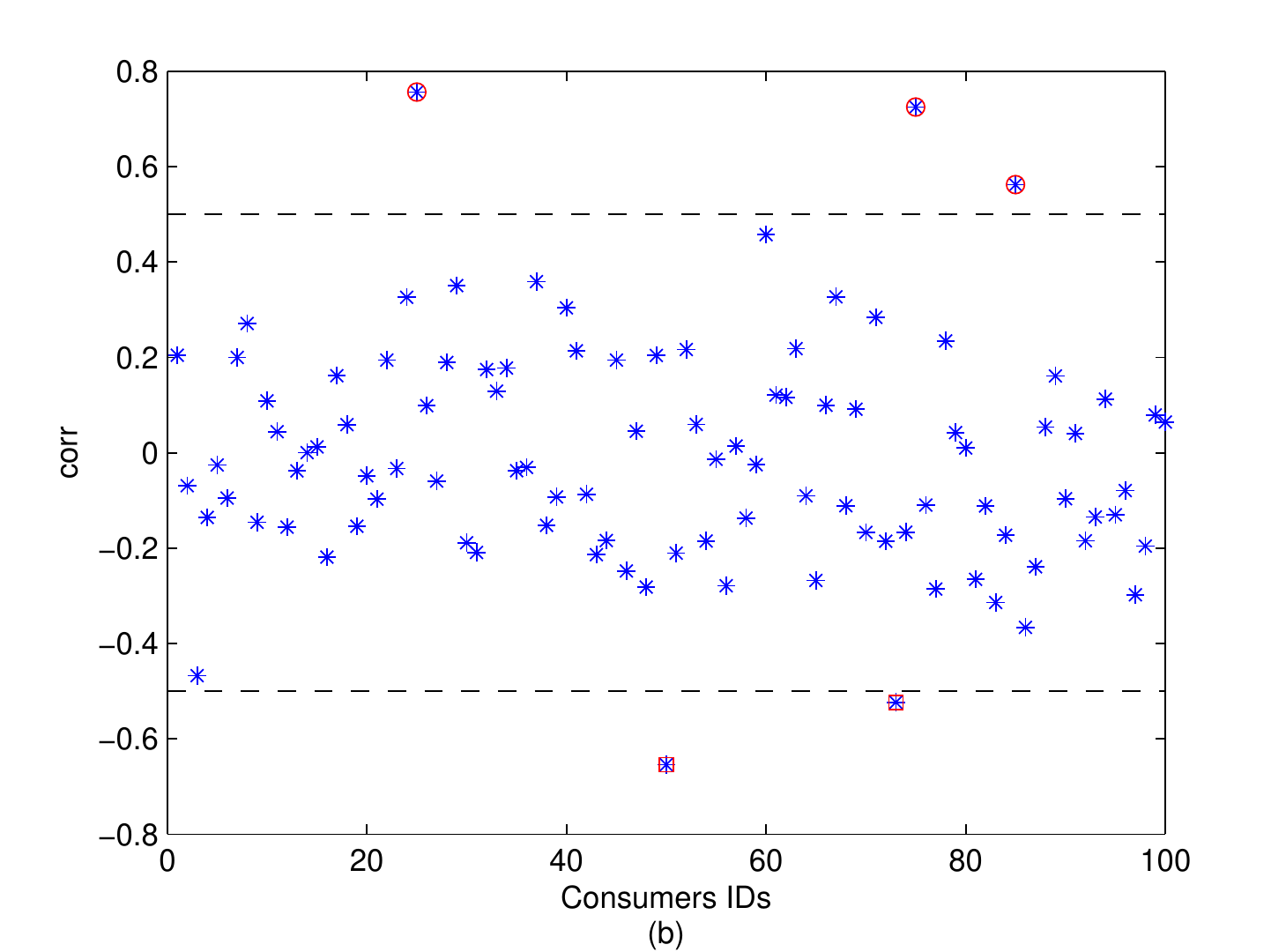}\\
		\includegraphics[width=\linewidth]{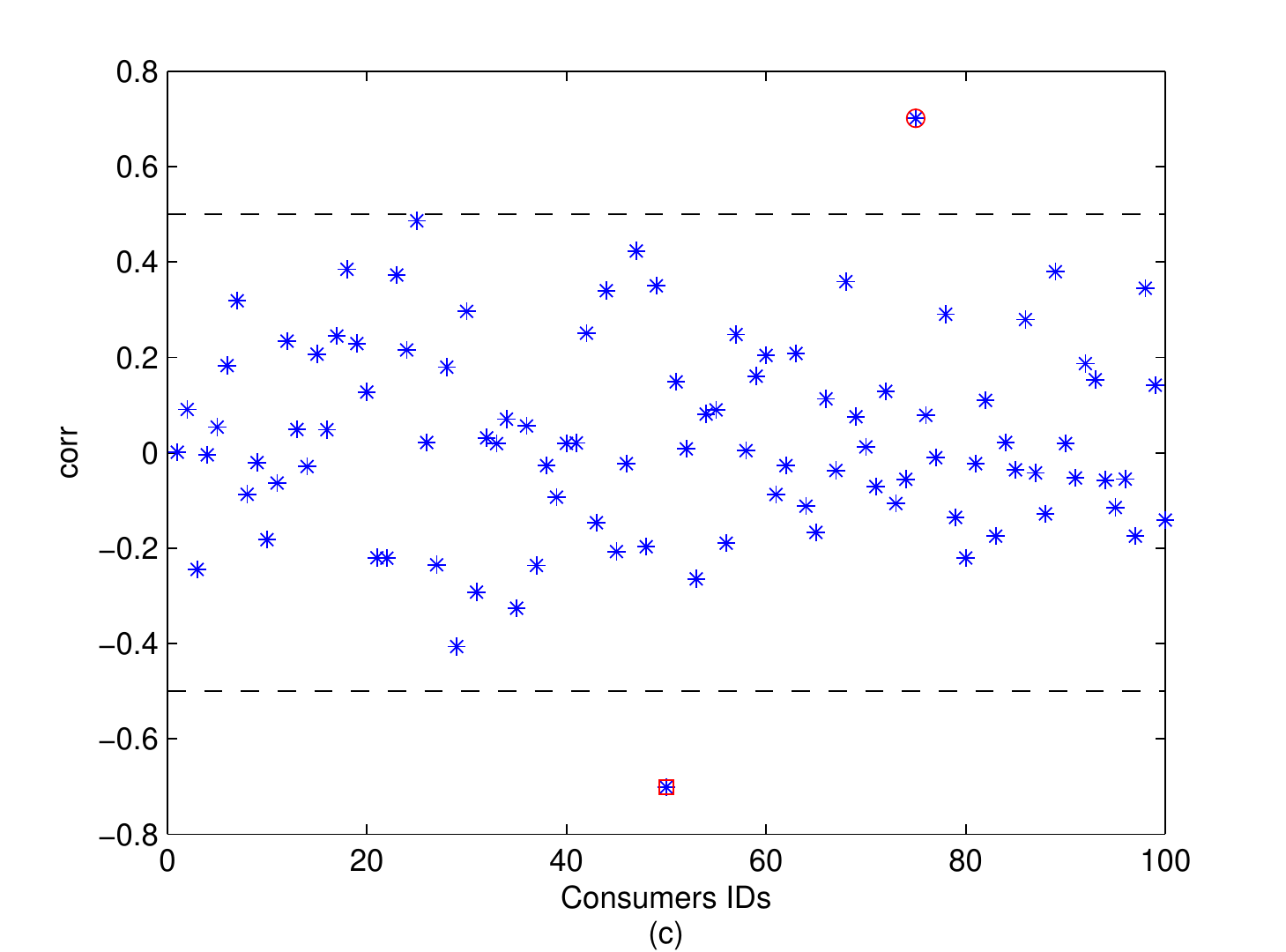}
	\end{center}
\caption{Existence of more than one malicious consumer. (a) all malicious consumer are detected correctly, (b) in addition to malicious consumers, a number of benign consumers are found malicious, and (c) not all malicious consumers are detected.}
\label{mixed}
\end{figure}

Next, we assume that there are three malicious consumers: $\# 25$, $\# 50$, and $\# 75$. Consumers with IDs $\# 25$ and $\# 75$ declare their power consumption less than their actual consumption and consumer $\# 50$ reports its power consumption more than its actual consumed energy. By setting the threshold to $0.5$, consumers with absolute value of correlation coefficient greater than $0.5$, i.e. $|corr|\geq 0.5$, would be considered malicious, as depicted in Fig.~\ref{mixed}.

As it can be seen from Fig.~\ref{mixed}, fixed threshold will result in three cases: 1) only malicious consumers been detected (Fig.~\ref{mixed}~(a)), 2) in addition to malicious consumers, some benign consumers found malicious (Fig.~\ref{mixed}~(b)), and 3) a subset of malicious consumers been detected (Fig.~\ref{mixed}~(c)).

\begin{figure}[t!]
	\begin{center}
		\includegraphics[width=\linewidth]{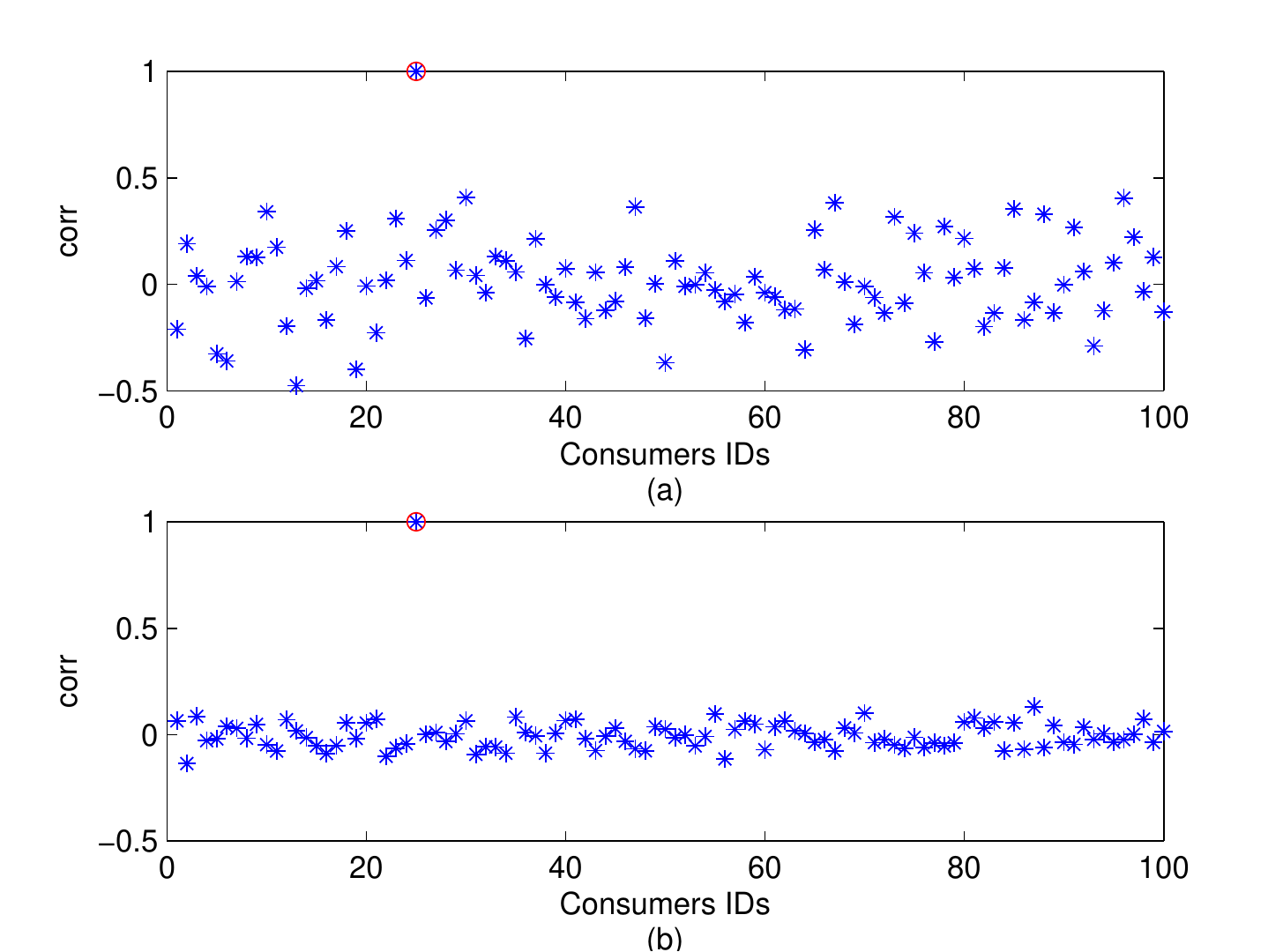}
		\caption{The effect of increasing in the number of samples on the detection rate. Fig.~\ref{12months} (a) and Fig.~\ref{12months} (b) illustrate the correlation coefficient of consumers within a month and a year respectively.}
		\label{12months}
	\end{center}
\end{figure}

As the proposed scheme is a statistical method, it is apparent that the more data be at hand, the more accurate the decision would be. Fig.~\ref{12months} illustrates this matter. In Fig.~\ref{12months}~(a), one month is considered as the period of measurement. On the other hand, the period of one year is considered in Fig.~\ref{12months}~(b). In this case, the number of samples has increased $12$ times. As a result, as it can be seen from Fig.~\ref{12months}~(b), the correlation coefficient corresponding to benign consumers revolves more densely around zero and the correlation coefficient corresponding to the malicious consumer (consumer $\# 25$) lies far apart from others compared to Fig.~\ref{12months} (a).

\subsection{Case II: Malicious consumer adds/subtracts a fixed quantity}
\begin{figure}[t!]
	\begin{center}
		\includegraphics[width=\linewidth]{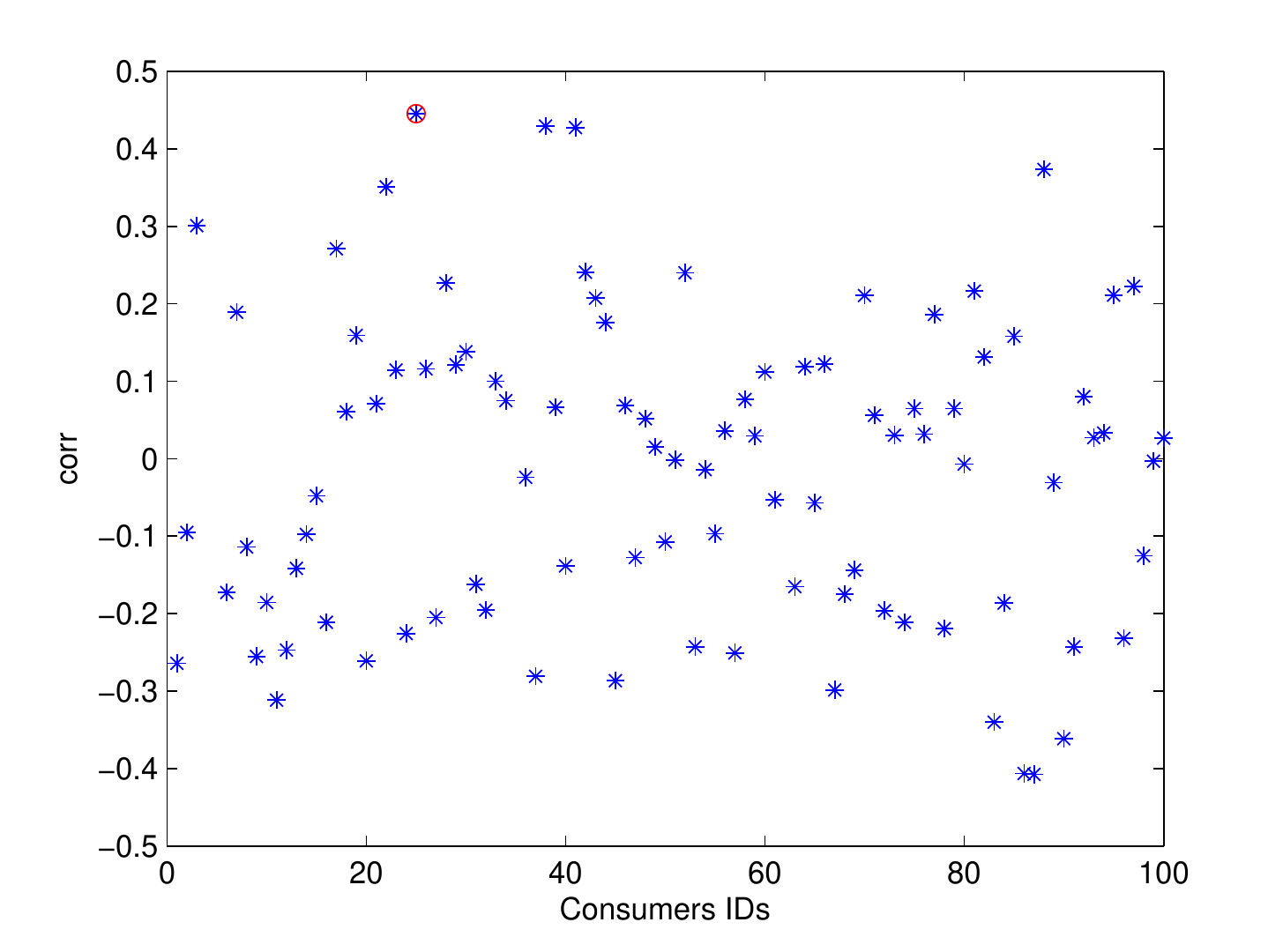}
		\caption{The correlation coefficient of the malicious consumer is  to $1$.}
		\label{case2-1}
	\end{center}
\end{figure}

Here we assume that the malicious consumers add/subtract the fixed quantity $\eta$ to/from the amount of consumed energy. This is independent from the actual usage of malicious consumer, detection of malicious consumer would be associated with those reported quantities that are equal to zero. This matter will result in the fact that detection of malicious consumers get more harder than the previous case, as illustrated in Fig.~\ref{case2-1}. More importantly, as it can be seen from Fig.~\ref{case2-1}, the correlation coefficient corresponding to the malicious consumer does not turn to $+1$. 

\begin{figure}[t!]
	\begin{center}
		\includegraphics[width=\linewidth]{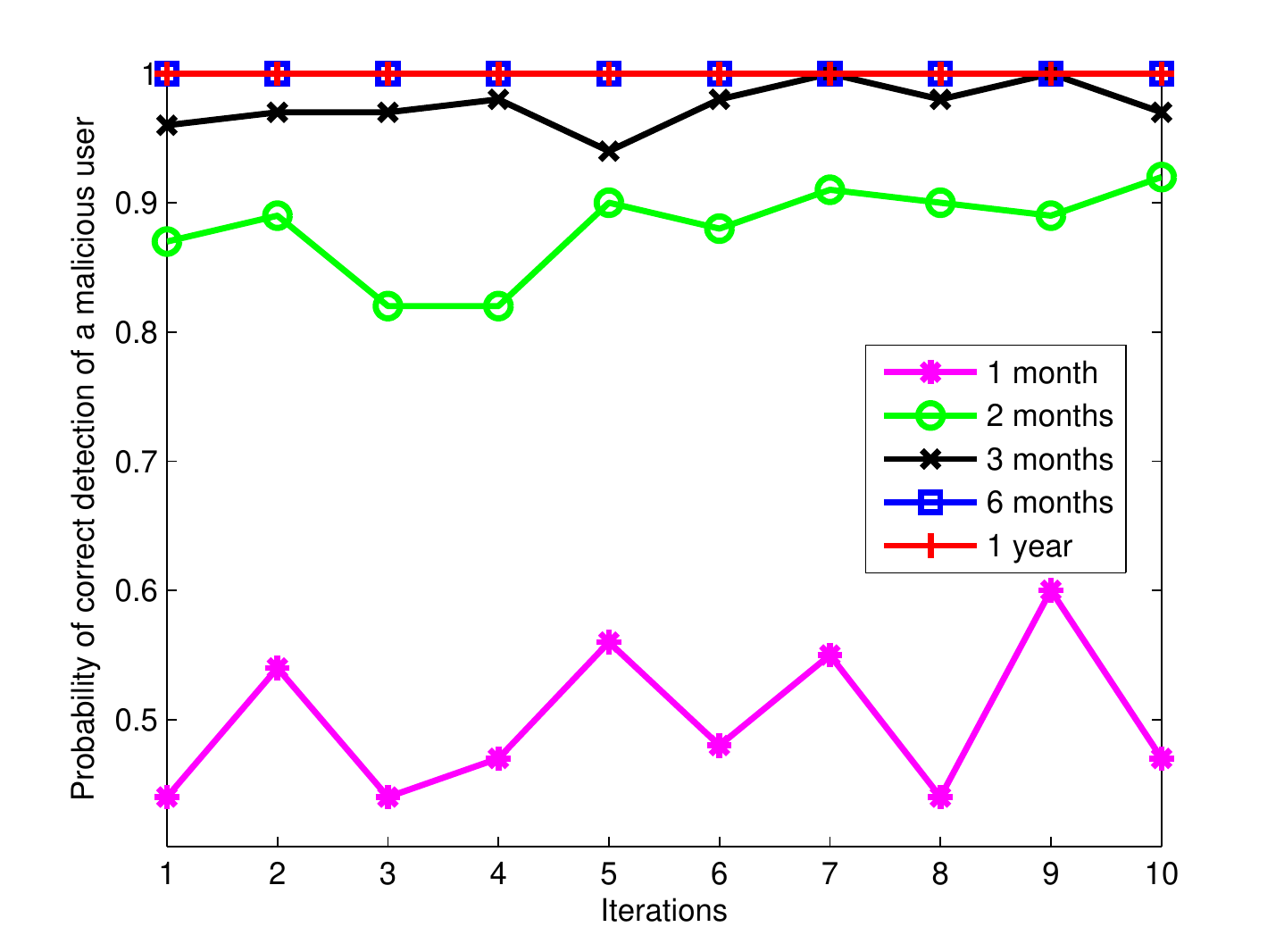}
		\caption{The probability of detection of malicious consumer improves as the duration of report analysis grow.}
		\label{case2_3}
	\end{center}
\end{figure}

It can be seen from Fig.~\ref{case2-1} that the correlation coefficient corresponding to the malicious consumer is far close to that of benign consumers. As stated before, this matter would made detection of the malicious consumer tough. Fig.~\ref{case2_3} illustrates this matter. As depicted in Fig.~\ref{case2_3}, the more samples we have corresponding to power consumption of consumers, the more accurate we are in detection of the malicious consumer. As it can be seen, by increasing the period of measurement, the probability of detecting the malicious consumer approaches to $1$, while for the period of one month this probability revolves around $0.5$.

In each iteration, the simulation is repeated $100$ times and the probability of correct detection of malicious consumer is calculated.

\subsection{Case III: Malicious consumer adds/subtracts a random quantity}
\begin{figure}[t!]
	\begin{center}
		\includegraphics[width=\linewidth]{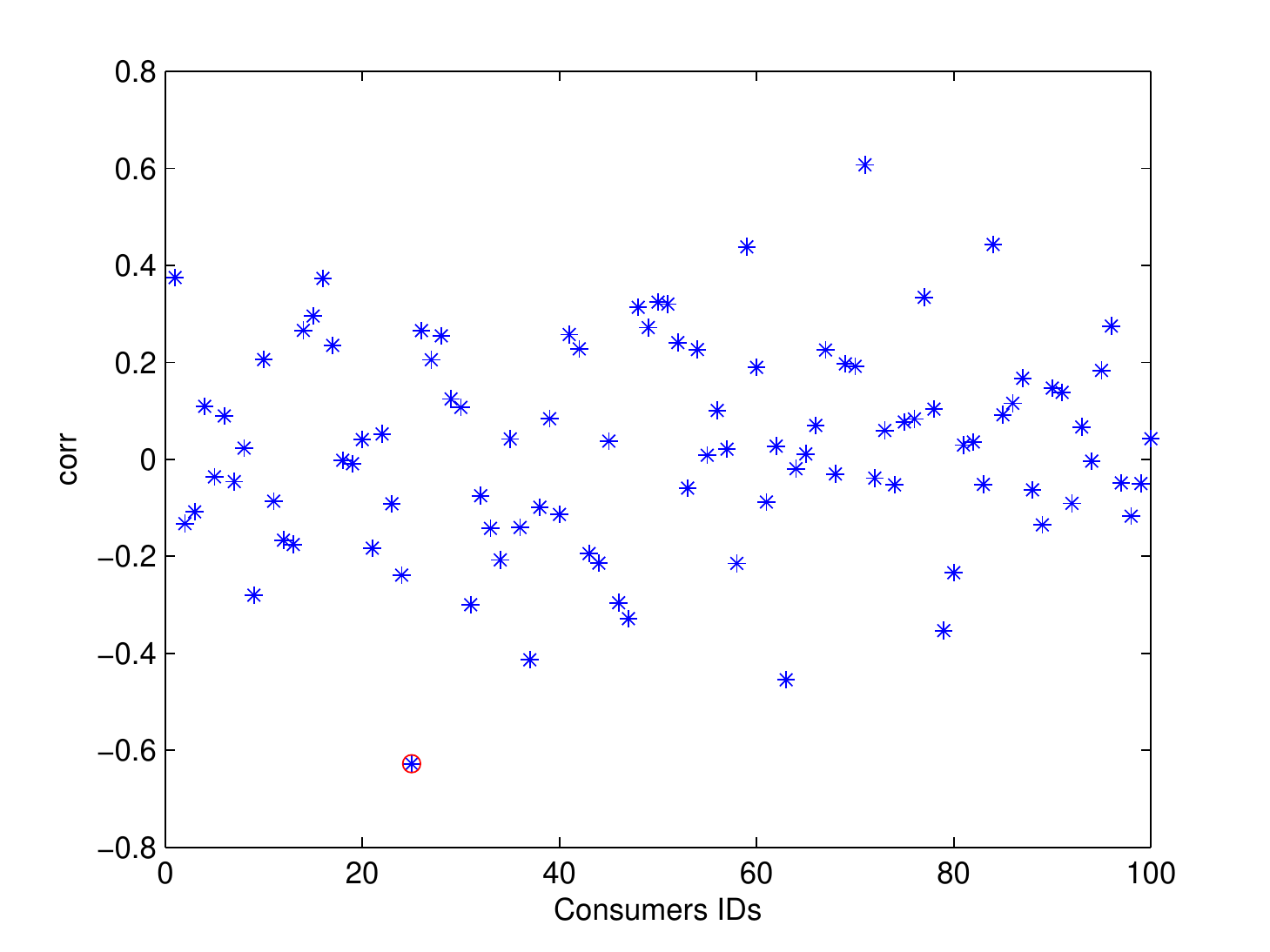}
		\caption{The correlation coefficient corresponding to benign and malicious consumers. The malicious consumer has the lowest correlation coefficient.}
		\label{case3_1}
	\end{center}
\end{figure}

In this section the simulation regarding the third scenario is brought where malicious consumer adds/subtracts a random quantity to its reported consumed energy.
As mentioned before, according to (\ref{case3}), the lowest negative quantity for the correlation coefficient expresses the malicious consumer, as depicted in Fig. \ref{case3_1}. Note that assumptions are similar to that of Fig.~\ref{Malicious} where consumer $\# 25$ is the malicious consumer unless mentioned otherwise.

The performance of our SBPP scheme in the three mentioned cases is evaluated in Table~\ref{Tbl.2}. Here, the simulation is repeated $1000$ times and the number of correct detections had been counted and finally the probability of correct detection had been calculated.
\begin{table} [t]
	\centering
	\caption{Probability of Correct Detection}
	\label{Tbl.2}
	\begin{tabular}{lcccc}
		\cline{2-5}
		& \multicolumn{4}{c}{Period of measurement}\\
		& 1 month & 3 months & 6 months & 1 year \\
		\hline
		Case I & 1 & 1 & 1 & 1 \\
		Case II & 0.51 & 0.98 & 1 & 1 \\
		Case III  & 0.92 & 1 & 1 & 1 \\
		\hline
	\end{tabular}
\end{table}

\section{Conclusion}
 \label{sec:conclusion}
We presented a statistical-based approach for data gathering in smart grids which preserves the privacy of consumers in the grid. We investigated the capability of the proposed scheme in detecting malicious consumers who dispatch bogus data to suppliers for a specific purpose such as abating their cost or imposing expenditure on them (subversive goals). Furthermore, we showed that when there exists at most one malicious consumer in each data gathering region, that consumer can be definitely detected. We consider three distinct cases for sabotage goals and show that our proposed method works properly in all cases. However, when the number of malicious consumers in a region grow, our statistical method would detect some benign consumers as malicious, or some malicious consumers remain undetected.

We then presented an algorithm for billing which concede the liability of billing to data aggregators in each region. By doing this, not only the signalling overhead decreases significantly, but also billing occurs at a trusted entity where malicious consumers are distinguished from benign ones. Our simulation results verified these terms.


\begin{thebibliography}{1}
	
	\bibitem{sbpp1}
	A. Ahadipour, M. Mohammadi, and A. Keshavarz-Haddad, \textquotedblleft SBPP: Statistical-Based Privacy-Preserving Scheme for Data Gathering in Smart Grid,\textquotedblright{} In \textit{OIC-CERT Journal of Cyber Security}, 2018.
	
	\bibitem{alabdulatif2017privacy}
	A. Alabdulatif, H. Kumarage, I. Khalil, M. Atiquzzaman, and X. Yi, \textquotedblleft Privacy-preserving cloud-based billing with lightweight homomorphic encryption for sensor-enabled smart grid infrastructure,\textquotedblright{} \textit{IET Wireless Sensor Systems}, vol. 7, no. 6, pp. 182-190, 2017.
	
	\bibitem{baloglu2018lightweight}
	U. B. BALOGLU and Y. DEM{\.I}R, \textquotedblleft Lightweight Privacy-Preserving Data Aggregation Scheme for Smart Grid Metering Infrastructure Protection,\textquotedblright{} \textit{International Journal of Critical Infrastructure Protection}, 2018.
	
	\bibitem{bao2015new}
	H. Bao and R. Lu, \textquotedblleft A new differentially private data aggregation with fault tolerance for smart grid communications,\textquotedblright{} \textit{IEEE Internet of Things Journal}, vol. 2, no. 3, pp. 248-258, 2015.
	
	\bibitem{boneh2005evaluating} 
	D. Boneh, E.-J. Goh, and K. Nissim, \textquotedblleft Evaluating 2-DNF formulas on ciphertexts,\textquotedblright{} In \textit{Theory of Cryptography Conference}, pp. 325-341, Springer, 2005.
	
	\bibitem{chen2015pdaft}
	L. Chen, R. Lu, and Z. Cao, \textquotedblleft PDAFT: A privacy-preserving data aggregation scheme with fault tolerance for smart grid communications,\textquotedblright{} \textit{Peer-to-peer networking and applications}, vol. 8, no. 6, pp. 1122-1132, 2015.
	
	\bibitem{chen2015muda}
	l. Chen, R. Lu, Z. Cao, K. AlHarbi, and X. Lin, \textquotedblleft MuDA: Multifunctional data aggregation in privacy-preserving smart grid communications,\textquotedblright{} \textit{Peer-to-peer networking and applications}, vol. 8, no. 5, pp. 777-792, 2015.
	
	\bibitem{chun2018privacy}
	H. Chun, K. Ren, W. and Jiang, \textquotedblleft Privacy-preserving power usage and supply control in smart grid,\textquotedblright{} \textit{Computers \& Security}, 2018.
	
	\bibitem{domingo2010coprivacy}
	J. Domingo-Ferrer, \textquotedblleft Coprivacy: towards a theory of sustainable privacy,\textquotedblright{} In \textit{International Conference on Privacy in Statistical Databases}, pp. 258-268, Springer, 2010.
	
	\bibitem{fadel2015survey}
	E. Fadel, V. C. Gungor, L. Nassef, N. Akkari, M. A. Malik, S. Almasri, and I. F. Akyildiz, \textquotedblleft A survey on wireless sensor networks for smart gridy,\textquotedblright{} \textit{Computer Communications}, vol. 71, pp. 22-33, 2015.
	
	\bibitem{fan2014privacy}
	C.-I. Fan, S.-Y. Huang, and Y.-L. Lai, \textquotedblleft Privacy-enhanced data aggregation scheme against internal attackers in smart grid,\textquotedblright{} \textit{IEEE Transactions on Industrial informatics}, vol. 10, no. 1, pp. 666-675, 2014. 
	
	\bibitem{ferrag1611survey}
	M. A. Ferrag, L. A. Maglaras, H. Janicke, and J. Jiang, \textquotedblleft A Survey on Privacy-preserving Schemes for Smart Grid Communications (2016),\textquotedblright{} \textit{arXiv preprint arXiv:1611.07722}, 2016.
	
	\bibitem{ferrer2011coprivacy}
	J. D. Ferrer, \textquotedblleft Coprivacy: an introduction to the theory and applications of co-operative privacy,\textquotedblright{} \textit{SORT: statistics and operations research transactions}, pp. 25-40, 2011.
	
	\bibitem{golle2008data}
	P. Golle, F. McSherry, and I. Mironov, \textquotedblleft Data collection with self-enforcing privacy,\textquotedblright{} \textit{ACM Transactions on Information and System Security (TISSEC)}, vol. 12, no. 2, pp. 9, 2008.
	
	\bibitem{he2016privacy}
	D. He, N. Kumar, and J.-H. Lee, \textquotedblleft Privacy-preserving data aggregation scheme against internal attackers in smart grids,\textquotedblright{} \textit{Wireless Networks}, vol. 22, no. 2, pp. 491-502, 2016.
	
	\bibitem{jia2014human}
	W. Jia, H. Zhu, Z. Cao, X. Dong, and C. Xiao, \textquotedblleft Human-factor-aware privacy-preserving aggregation in smart grid,\textquotedblright{} \textit{IEEE Systems Journal}, vol. 8, no. 2, pp. 598-607, 2014.
	
	\bibitem{jiang2018achieving}
	R. Jiang, R. Lu, and K.-K. R. Choo, \textquotedblleft Achieving high performance and privacy-preserving query over encrypted multidimensional big metering data,\textquotedblright{} \textit{Future Generation Computer Systems}, vol. 78, pp. 392-401, 2018.
	
	\bibitem{kumar2010freedom}
	R. Kumar, R. Gopal, and R. Garfinkel, \textquotedblleft Freedom of privacy: anonymous data collection with respondent-defined privacy protection,\textquotedblright{} \textit{INFORMS Journal on Computing}, vol. 22, no. 3, pp.471-481, 2010,
	
	\bibitem{li2015pda}
	C. Li, R. Lu, H. Li, L. Chen, Le and J. Chen, \textquotedblleft PDA: a privacy-preserving dual-functional aggregation scheme for smart grid communications,\textquotedblright{} \textit{Security and Communication Networks}, vol. 8, no. 15, pp. 2494-2506, 2015.
	
	\bibitem{li2014enabling}
	H. Li, \textquotedblleft Enabling Secure and Privacy Preserving Communications in Smart Grids,\textquotedblright{} Springer, 2014.
	
	\bibitem{liao2017optimal}
	G. Liao, X. Chen, and J. Huang, \textquotedblleft Optimal Privacy-Preserving Data Collection: A Prospect Theory Perspective,\textquotedblright{} In \textit{GLOBECOM 2017-2017 IEEE Global Communications Conference}, pp.1-6, IEEE, 2017.
	
	\bibitem{lu2016privacy}
	R. Lu, \textquotedblleft Privacy-enhancing aggregation techniques for smart grid communications,\textquotedblright{} Springer, 2016.
	
	\bibitem{paillier1999public}
	P. Paillier, \textquotedblleft Public-key cryptosystems based on composite degree residuosity classes,\textquotedblright{} In \textit{International Conference on the Theory and Applications of Cryptographic Techniques}, pp. 223-238, Springer, 1999.
	
	\bibitem{rial2018privacy}
	A. Rial, G. Danezis, and M. Kohlweiss, \textquotedblleft Privacy-preserving smart metering revisited,\textquotedblright{} \textit{International Journal of Information Security}, vol. 17, no. 1, pp. 1-31, 2018.
	
	\bibitem{rottondi2013distributed}
	C. Rottondi, G. Verticale, and C. Krauss, \textquotedblleft Distributed privacy-preserving aggregation of metering data in smart grids,\textquotedblright{} \textit{IEEE Journal on Selected Areas in Communications}, vol. 31, no. 7, pp. 1342-1354, 2013.
	
	\bibitem{shi2015diverse}
	Z. Shi, R. Sun, R. Lu, L. Chen, J. Chen, and X. S. Shen, \textquotedblleft Diverse grouping-based aggregation protocol with error detection for smart grid communications,\textquotedblright{} \textit{IEEE Transactions on Smart Grid}, vol. 6, no. 6, pp. 2856-2868, 2015.
	
	\bibitem{csimcsek2018tps3}
	M. U. {\c{S}}im{\c{s}}ek, F. Y{\i}ld{\i}r{\i}m Okay, D. Mert, and S. {\"O}zdemir, \textquotedblleft TPS3: A privacy preserving data collection protocol for smart grids,\textquotedblright{} \textit{Information Security Journal: A Global Perspective}, vol. 27, no. 2, pp. 102-118, 2018.
	
	\bibitem{stegelmann2010towards}
	M. Stegelmann, \textquotedblleft Towards fair indictment for data collection with self-enforcing privacy,\textquotedblright{} In \textit{IFIP International Information Security Conference}, pp. 265-276, Springer, 2010.
	
	\bibitem{sun2013aped}
	R. Sun, Z. Shi, R. Lu, M. Lu, Min and X. S. Shen, \textquotedblleft APED: An efficient aggregation protocol with error detection for smart grid communications,\textquotedblright{} In \textit{Global Communications Conference (GLOBECOM), 2013 IEEE}, pp. 432-437, IEEE, 2013.
	
	\bibitem{wang2016balanced}
	H. Wang, D. He, and S. Zhang, \textquotedblleft Balanced anonymity and traceability for outsourcing small-scale data linear aggregation in the smart grid,\textquotedblright{} \textit{IET Information Security}, vol. 11, no. 3, pp. 131-138, 2016.
	
	\bibitem{wang2015tpp}
	H. Wang, B. Qin, Q. Wu, L. Xu, and J. Domingo-Ferrer, \textquotedblleft TPP: Traceable privacy-preserving communication and precise reward for vehicle-to-grid networks in smart grids,\textquotedblright{} \textit{IEEE Transactions on Information Forensics and Security}, vol. 10, no. 11, pp. 2340-2351, 2015.
	
	\bibitem{wong2014privacy}
	K.-S. Wong and M. H. Kim, \textquotedblleft Privacy-preserving data collection with self-awareness protection,\textquotedblright{} In \textit{Frontier and Innovation in Future Computing and Communications}, pp. 365-371, Springer, 2014.
	
	\bibitem{XiaXLZ2018}
	X. Xia, Y. Xiao, W. Liang, and M. Zheng, \textquotedblleft GTHI: A Heuristic Algorithm to Detect Malicious Users in Smart Grids,\textquotedblright{} \textit{IEEE Transactions on Network Science and Engineering}, 2018.
	
	\bibitem{xu2018privacy}
	L. Xu, C. Jiang, Y. Qian, Y. Ren, L. Xu, C. Jiang, Y. Qian, and Y. Ren, \textquotedblleft Privacy-Preserving Data Collecting: A Simple Game Theoretic Approach,\textquotedblright{} \textit{Data Privacy Games}, pp. 45-57, 2018.
	
	
\end{thebibliography}
\end{document}